\documentclass[prl,twocolumn, runinaddress,preprintnumbers,showpacs,a4paper]{revtex4}

\usepackage{graphicx}

\usepackage{amsmath}
\usepackage{amssymb}
\usepackage{amstext}
\usepackage{amsxtra}
\usepackage{version}
\begin{document}
%macros
\newcommand{\Tc}{T$_{\mathrm C}$}
\newcommand{\Tco }{T$_{\mathrm CO}$\space}
\newcommand{\sinth}{\mbox{$\sin\theta/\lambda$}}
\newcommand{\inA}{\mbox{\AA$^{-1}$}}
\newcommand{\mub}{\mbox{$\mu_{B}$}}
\newcommand{\mns}{$-$}
\newcommand{\ddd}{3\textit{d}}
\newcommand{\pp}{2\textit{p}}
\newlength{\minusspace}
\settowidth{\minusspace}{$-$}
\newcommand{\msp}{\hspace*{\minusspace}}
\newlength{\zerospace}
\settowidth{\zerospace}{$0$}
\newcommand{\zsp}{\hspace*{\zerospace}}
\newcommand{\lasrmn}{La$_{2-2x}$Sr$_{1+2x}$Mn$_{2}$O$_7$\space}
\newcommand{\mnt}{$Mn^{3+}$}
\newcommand{\mnf}{$Mn^{4+}$}
\newcommand{\eg}{$\textit{e}_{g}$ }
\newcommand{\qce}{$\textit{q}_{CE}=(1/4,1/4,0)$}
\newcommand{\qcel}{$\textit{q}_{CE}=(1/4,-1/4,0)$}
\newcommand{\qos}{$\textit{q}_{L}=(0.3,0,1)$}
\newcommand{\sqw}{S(\textit{Q},$\omega$)}
\newcommand{\Ts}{T$^{*}$\space}
\newcommand{\GG}{$\Gamma$\space}
\newcommand{\NaH}{Na$_{x}$CoO$_{2}\cdot y$H$_{2}$O}
\newcommand{\NaD}{Na$_{x}$CoO$_{2}\cdot y$D$_{2}$O}
\newcommand{\NaCo}{Na$_{x}$CoO$_{2}$}
\newcommand{\wat}{H$_{2}$O}
\newcommand{\deut}{D$_{2}$O}
\newcommand{\etal}{\textit{et al.}\space}
\newcommand{\degc}{$^{\circ}$C\space}
\newcommand{\br}{Br$_{2}$\space}
\newcommand{\xrange}{$0.28<x<0.37$}
\newcommand{\nrange}{$0.24<n<0.35$}
\newcommand{\ox}{H$_{3}$O}
\newcommand{\oxp}{H$_{3}$O$^{+}$}
\newcommand{\NaHox}{Na$_{x}$(H$_{3}$O)$_{z}$CoO$_{2}\cdot y$H$_{2}$O}
\newcommand{\NaHoxp}{Na$_{x}^{+}$(H$_{3}$O)$_{z}^{+}$CoO$_{2}\cdot y$H$_{2}$O}


\title{Revised superconducting phase diagram of hole doped \NaHox}
% repeat the \author .. \affiliation  e\Tc\space . as needed

% other informatio
\author{C. J. Milne}
\affiliation{Hahn-Meitner-Institut, Glienicker Str. 100, Berlin D-14109, Germany}
\author{D. N. Argyriou}
\email[Email of corresponding author: ]{argyriou@hmi.de}
\affiliation{Hahn-Meitner-Institut, Glienicker Str. 100, Berlin D-14109, Germany}
\author{A. Chemseddine}
\affiliation{Hahn-Meitner-Institut, Glienicker Str. 100, Berlin D-14109, Germany}
\author{N. Aliouane}
\affiliation{Hahn-Meitner-Institut, Glienicker Str. 100, Berlin D-14109, Germany}
\author{J. Veira}
\affiliation{Hahn-Meitner-Institut, Glienicker Str. 100, Berlin D-14109, Germany}
\author{S. Landsgesell}
\affiliation{Hahn-Meitner-Institut, Glienicker Str. 100, Berlin D-14109, Germany}
\author{D. Alber}
\affiliation{Hahn-Meitner-Institut, Glienicker Str. 100, Berlin D-14109, Germany}

%\date{\today}
%\preprint{Revised 28 October, 2004}
\begin{abstract}
We have studied the superconducting phase diagram of \NaH\space as a function of electronic doping, characterizing our samples both in terms of Na content $x$ and the Co valence state. Our findings are consistent with a recent report that intercalation of \oxp\space ions into Na$_{x}$CoO$_{2}$, together  with water, act as an additional dopant indicating that Na sub-stochiometry alone does not control the electronic doping of these materials. We find a superconducting phase diagram where optimal \Tc\space is achieved through a Co valence range of 3.24 - 3.35, while \Tc\space decreases for materials with a higher Co valence.  The critical role of dimensionality in achieving  superconductivity is highlighted by similarly doped non-superconducting anhydrous samples, differing from the superconducting hydrate only in inter-layer spacing. The increase of the interlayer separation between CoO$_{2}$ sheets as Co valence is varied into the optimal \Tc\space region is further evidence for this criticality.
\end{abstract}
\pacs{74.25.Dw, 74.70.-b}
%\keywords{}

\maketitle

A characteristic feature of the cuprate superconductors is the existence of an optimal electronic doping that gives a maximum superconducting transition temperature, \Tc. This composition separates the under-doped and over-doped regimes in which \Tc\space decreases from the optimal value.  This behavior is thought to be a universal characteristic of cuprate superconductors that arises from the fundamental origin of superconductivity in these systems. This feature of the cuprates has recently been tested for the layered intercalated cobaltate  \NaH (x$\sim$0.35 and y$\sim$1.3),\cite{Schaak} that exhibits superconductivity below 5K.\cite{Takada} The structure of these layered cobaltates is significantly different from the cuprates, in that superconductivity is thought to occurs in  CoO$_{2}$ sheets that have quasi 2D triangular symmetry, analogous to that of geometrically frustrated systems.  An initial superconducting phase diagram of \NaCo$\cdot$1.3\wat, was drawn as a function of Na content $x$.\cite{Schaak} Schaak \etal demonstrated that such a phase diagram resembles the characteristic phase diagram of the cuprates in that an optimum  \Tc=4.5K is found for a material of Na content $x$= 0.3, with \Tc\space rapidly decreasing for lower or higher Na contents.\cite{Schaak} 

The stoichiometry of this superconductor has recently been revised by Takada \etal\cite{Takada2}, to account for a lower Co valence than that expected if the system is doped purely by Na non-stoichiometry.  This work showed that charge neutrality is reached by the intercalation of oxonium ions (\oxp) along with water, giving a composition of this superconducting phase of Na$_{0.337}$(\ox)$_{0.234}$CoO$_{2}\cdot y$\wat.\cite{Takada2}  Our examination of the superconducting phase diagram of \NaHox, where we determine the Co valence via redox titrations, demonstrates that the optimal \Tc\space for this superconductor is obtained over the wide Co valence range, 3.24 - 3.35, while \Tc\space decreases for samples with a Co valence $>$ 3.35. 
Although X-ray absorption and photoemission spectroscopy\cite{kubota, chainani} have been applied to probe the electronic states of both the anhydrous and hydrated layered cobaltates, and identified Co$^{3+}$ and Co$^{4+}$ species, quantitative data is not attainable. In the absence of ARPES data, to map the Fermi surface, simple titrations are an accurate and reliable technique to determine the Co valence. Similar methods have been applied to determine the valence of Cu in the cuprates.\cite{Manthiram}

The fundamental electronic and magnetic interactions in the parent compounds of the superconducting phase are tuned by the doping of 1-$x$ charge carriers on formation of sub-stoichiometric Na$_{x}$CoO$_{2}$. In the stoichiometric $x$= 1 compound Co is in the Co$^{3+}$ state with $S$=0 in a low spin (LS) $t_{2g}^{6}$  configuration. For $x<$1 the result is a mixed valence system Na$_{x}$Co$^{3+}_{x}$Co$^{4+}_{1-x}$O$_{2}$ where Co$^{4+}$ has a LS $S = $1/2 configuration with holes doped into a $t_{2g}^{5}$  state. This provides control for both the electronic and magnetic properties of these materials in a natural way, where Na content $x$ controls both the charge carrier concentration and the amount of magnetic LS Co$^{4+}$ ions. For example the $x=$0.75 compound has attracted attention due to its unusual thermoelectric properties,\cite{Motohashi1, Motohashi2,Tojo,Wang} and it has recently been viewed as a correlated electron system with correlations driving coincident magnetic and electronic transitions.

We have synthesized a series of samples with varying Na content $x$, and have chemically characterized them in terms of both $x$, using neutron activation analysis, and the Co valence, using redox titrations. This has allowed us to accurately determine the parameters that control electronic doping in these materials and draw the superconducting phase diagram in terms of Co valence. Our measurements indicate that the  amount of \oxp\space that is intercalated into the lattice has an  approximate inverse relationship with respect to the Na content $x$. We find optimum \Tc\space is achieved over the cobalt valence range of 3.24 - 3.35, while \Tc\space decreases for valence states $>$ 3.35. Measurements of the $c/a$ ratio of the lattice constants indicate that the separation between  CoO$_{2}$ sheets increases as the Co valence is tuned into the optimal region. This suggests that optimum \Tc\space is achieved  as  the electronic interlayer coupling between CoO$_{2}$ sheets becomes more 2D-like.

Polycrystalline samples of Na$_{x}$CoO$_{2}$, $x$$\sim$0.70, were first prepared by mixing appropriate amounts  Na$_{2}$CO$_{3}$ and Co$_{3}$O$_{4}$  with a 10\% molar excess of Na. These materials were heated to 850\degc in flowing oxygen for 36 hours with 2 intermediate regrindings. This method yielded x-ray pure phases of Na$_{0.72}$CoO$_{2}$ (for determination of $x$ see below). In order to move from $x$=0.72 to Na contents where superconductivity is observed, powder samples were immersed in a bromine-acetonitrile solution with a  2, 5, 36, and 50  times bromine excess as compared to Na. Samples were stirred in the solution for 24 hours  after which they were washed with acetonitrile and then \wat\space or D$_{2}$O. A time dependence study of the de-intercalation process showed that it reaches completion within a few hours as opposed to 5 days as suggested previously.\cite{Schaak,Takada}. The Na/Co ratio of the products was measured using neutron activation analysis (NAA) to establish $x$.
\footnote{NAA provides quantitative analysis of the chemical composition of a sample by measuring the decay of characteristic $\gamma$-rays due to neutron capture on irradiation with neutrons inside a nuclear reactor.} To achieve superconductivity the Na de-intercalated materials were placed in a sealed container with either \wat\space or D$_{2}$O and evacuated to a pressure of $\sim$ 20-30 mbar. Sample containers were kept at a constant temperature of 26\degc to produce a $p(H_{2}O)$ of $\sim$ 40 mbar. Samples were left in a humid atmosphere for at least two weeks before final characterization measurements were made. 

\begin{figure}[ !t]
\includegraphics[width=0.4\textwidth, angle=-90]{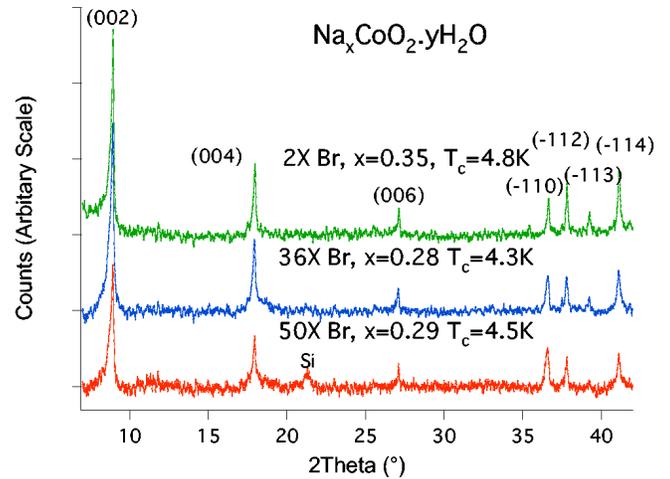}
\caption{XRD measurements of three hydrated samples prepared for this work. These measurements are typical of all the sample prepared in this work. The Si peak from the sample holder is indicated for the last diffraction pattern.}
\label{xrd1}
\end{figure}

The valence of Co was measured directly by redox titration as described in ref. \onlinecite{Takada2}. As a check of the accuracy of this method, the Co valence of an anhydrous sample of Na$_{0.66(1)}$CoO$_{2}$ was measured to be 3.35, in excellent agreement with the $4-x$ value determined from NAA. The error in the measurement of the Co valence is estimated to be 0.01. The crystal structure of these materials was probed using a Guinear X-ray diffractometer (XRD) with Cu$K\alpha$ source which confirmed that all samples crystallized in a hexagonal structure as described by Takada \etal \cite{Takada} and free of intermediate hydrate phases.  Typical diffraction patterns from these samples are shown in fig.~\ref{xrd1}. Rietveld analysis of the XRD measurements was used to determine lattice constants for our samples. The Na/Co ratio of each sample was determined using NAA. The error in $x$ from these measurements is estimated to be 5\%.  Superconducting properties were characterized using a Quantum design SQUID magnetometer. In table~\ref{results} we show the summary of our results from the  characterization of our samples in terms of Na content $x$, Co valence, lattice constants and superconducting transition temperatures. 

\begin{figure}[! tbp]
\includegraphics[scale=0.30,angle=-90]{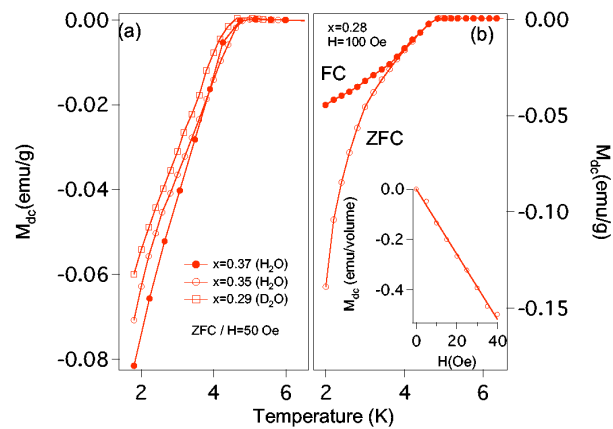}
\caption{Typical dc magnetization measurements of samples investigated in this work. (a) ZFC dc magnetization measurements in H=50 Oe.  (b) FC and ZFC measurements in H=100 Oe for the $x$=0.28 (\deut) sample. The inset shows a M vs H measured from the $x$=0.35(\deut) sample at 1.8K. The data without corrections for pinning, shape or excess water indicates a superconducting volume fraction of $\sim$20\%.}
\label{chi}
\end{figure}

\begingroup
\squeezetable
\begin{table}[!b]
\caption{Details of the preparation, hydration medium and characterization of the samples used in this work. The oxidation state of Co was determined for most samples used in this work from redox titrations. The superconducting transition temperatures, \Tc,  were determined using a SQUID magnetometer defining \Tc\space as the onset of diamagnetism in ZFC and H= 50 Oe.}
\label{results}
\begin{ruledtabular}
%\begin{tabular}{lllll}
\begin{tabular}{cccccccc}
Br\\ Excess & $x$  &Co$^{\alpha}$& \wat/\deut & $a$ ($\AA$)& $c$($\AA$)&$c/a$&\Tc\space \\
\hline
2&0.41&3.38&\deut&2.8278(14)&19.561(2)	&6.917(10)&2.8\\%xtal2x2
2&0.37&-&\wat&2.8249(8)&19.657(2)&6.959(6)&4.5	\\%	JAN2_2X
5&0.37&3.45& (H,D)$_{2}$O&2.8257(8)&	19.578(7)&6.928(6	)&2.5\\%xtal2x5b
5&0.35&3.32&\deut&2.8244(7)&19.656(6)&	6.959(5)&4.8\\%xtal2x5
5&0.35&3.33&\wat&2.8231(6)&19.668(2)&6.967(4)&4.8\\%JAN2_5X_Vac
5&0.35&3.24&\deut&2.8253(11	)&19.676(8)&6.971(8)&4.7\\%	JAN4_5X5
36&0.32&3.27&\deut&2.8231(10)&19.784(7	)&7.008(7)&	4.5\\%CM100D
36&0.28&-&\wat&2.8251(9)&19.677(7)&6.965(7)&4.3\\%JAN2_36X_Vac
36&0.28&3.35&\deut&2.8245(5)&19.726(9)	&6.975(5)&4.3\\%JAN4_5X36
50&0.29&-&\deut&2.8261(7)&19.653(7)&6.954(6)&4.5\\%Jan2_50X_Vac
50&0.29&3.26&\deut&2.8233(4)&19.855(7)	&7.033(4)&4.4\\%xtal2x50
\end{tabular}
\end{ruledtabular}
\end{table}
\endgroup

As suggested by the results shown in table ~\ref{results}, there is a significant discrepancy between the expected valence of Co based purely on Na content and that given by redox titration.  These observations are consistent with the revised composition of the superconducting phase 
where \oxp\space accounts for the missing charge. From our measurements of $x$ and the Co valence we can determine the amount of \oxp\space present in our samples. We find that there is a relationship between $x$ and the Co valence, indicating that Na poor samples tend to contain larger amounts of \oxp\space(see inset in fig. 3b). This is consistent with recent structural measurements suggesting that Na$^{+}$ and \oxp\space may reside on the same crystallographic site.\cite{Takada2} Our results show that as Na is de-intercalated from the lattice to values of $x<\frac{1}{2}$, a cobalt valence $>$3.5 is achieved, the intercalation of \oxp\space however, reduces the cobalt valence  to values of $<$3.5.\cite{Takada2} To determine the properties of this novel superconductor as a function of electronic doping, we have produced a series of samples with varying cobalt valence. By controlling the amount of Na chemically, and measuring directly the Co valence we have drawn a phase diagram using a series of samples over the broad range of cobalt valence $3.24 - 3.45$. 

Typical magnetization measurements of our samples are shown in fig.~\ref{chi}a. Here the zero field cooling (ZFC) dc magnetizations at 1.8K are in the range of M$_{dc}$=0.7-1.6$\times10^{-3}$(emu/g/Oe) very similar to those obtained by other workers for similar \Tc\space samples (1.8-5.5$\times10^{-3}$(emu/g/Oe)).\cite{Takada,Schaak} Estimation of superconducting volume fractions was performed using both field cooling (FC) and ZFC measurements as well as measurements of M vs H (see fig. ~\ref{chi}b). Overall we found values of superconducting volume fraction to be in the range of 20 to 30\%, indicative of bulk superconductivity for samples with cobalt valence states between $3.24$ and $3.35$. However, these values are only estimates as there is a significant error in the measurement of the mass of the samples due to unbound inter-granular water.  This would tend to underestimate the superconducting volume fraction.  Overall the \Tc\space values obtained in this work are relatively high ($\ge$ 4.3K) with the exception of two samples (see table ~\ref{results}) with \Tc=2.5 and 2.8K. These two samples exhibited a low diamagnetic signal and \Tc\space was measured using magnetic susceptibility measurements. In agreement with the measurements of Jin \etal \cite{Jin} no clear difference is found between the \Tc\space of hydrated and deuterated samples. Finally we comment that the diamagnetic transitions of all samples are relatively broad in temperature,  suggesting an appreciable degree of inhomogeneity.  This behavior appears to be typical of these materials\cite{Jin,Schaak} and is understandable in view of the synthesis method of these superconductors.

\begin{figure}[ !t]
\includegraphics[scale=0.32]{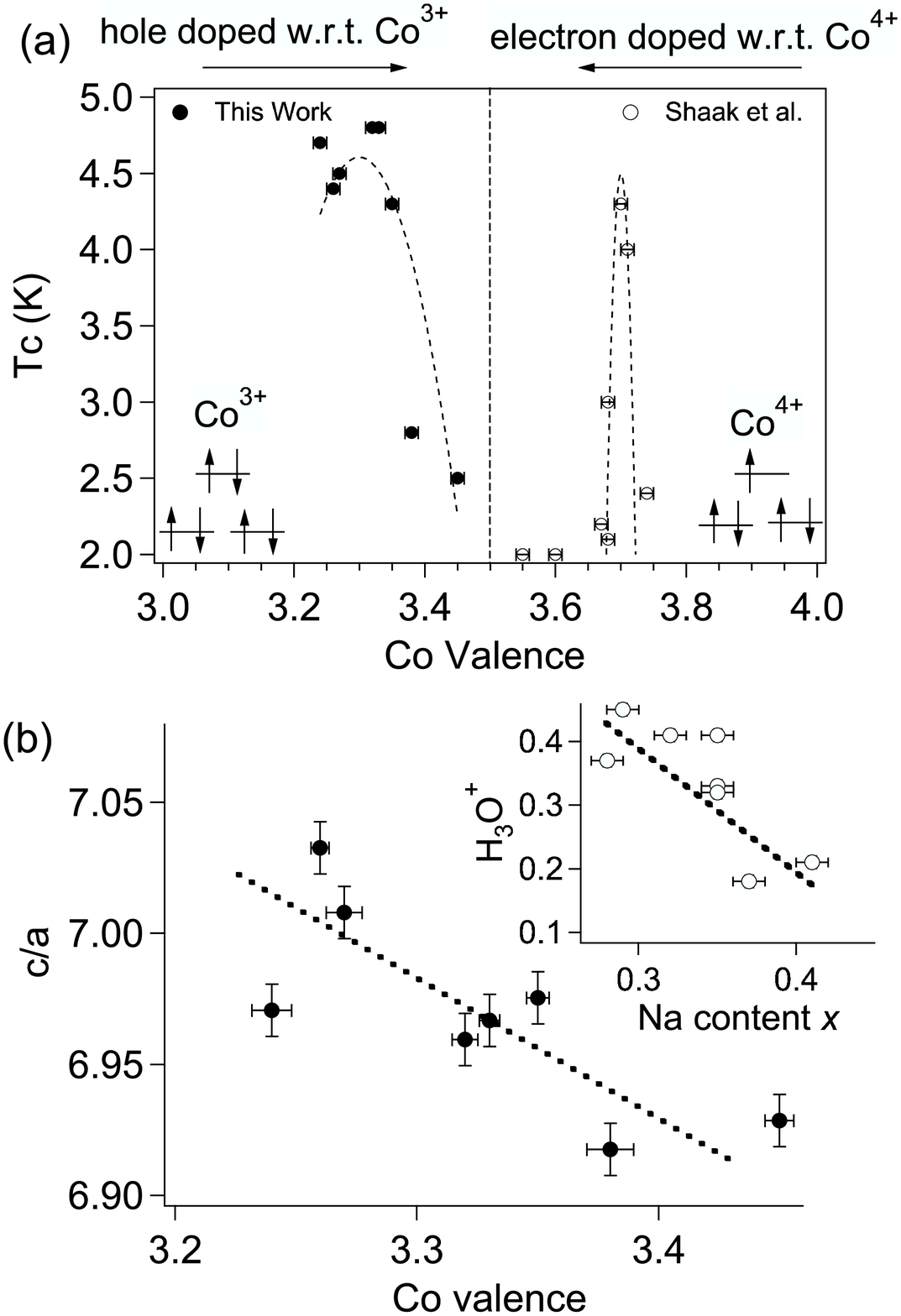}
\caption{(a)\Tc\space as a function of the Co valence state given in table ~\ref{results}. The left curve was obtained from our titration data whilst the right hand curve describes oxidation states based solely on Na content from Schaak et al.\cite{Schaak} (b) The variation of the $c/a$ ratio as a function of cobalt oxidation state. In the insert we show the variation of the \oxp content of our samples as determined from measurements of the Na content and the Co oxidation state. Dashed lines are a guide to the eye. }
\label{tcvsx}
\end{figure}

The superconducting phase diagram of \NaHox\space determined from our measurements is shown in fig. ~\ref{tcvsx}a. From these data we find that optimal \Tc\space is reached over the region of cobalt valence $3.24 - 3.35$, while for cobalt valence states $>3.35$ \Tc\space decreases to values of $<$ 3K. In the optimal cobalt valence region \Tc\space differs from 4.3 to 4.8 K values close to the maximum values of \Tc\space reported for this material.\cite{Takada}    

The cobalt valence is a reflection of the electronic doping in the material. We can consider the cobalt valence range 3.0 - 3.5 as hole doped with respect to Co$^{3+}$ , with the number of holes, $n$, equal to (valence state)-3. Holes are doped in by removal of electrons from the upper lying, fully occupied, two electron t$_{2}$g band of Co in trigonally distorted CoO$_{2}$ sheets (see fig. 3a). Conversely, the cobalt valence range $3.5 - 4.0$ can be considered as electron doped with respect to Co$^{4+}$, with the number of electrons determined as 4 -(valence state), with doped electrons entering the upper lying, half filled, two electron t$_{2}$g band. 

The phase diagram that we draw here, by measuring the Co valence directly, demonstrates that superconductivity in these layered cobaltates occurs for hole doped compositions, and is in sharp contrast to the phase diagram of Shaak et al.\cite{Schaak}, where the Co valence was thought to be controlled by Na-content alone. The width of the superconducting region can not be defined at this stage as we have not been able to obtain underdoped samples, however from the present data we estimate it to be $\Delta n\sim$0.15-0.2, a value actually similar to that found for hole doped cuprate superconductors such as for example in La$_{1-x}$Sr$_{x}$CuO$_{4-\delta}$.\cite{Radaelli} Although our data suggests similarities with the band filling model of the cuprates, superconductivity in this system is observed on hole doping of a fully occupied band, as opposed to the hole doping of a half filled band as found in the cuprates. 

In fig. ~\ref{tcvsx}b we show the variation of the ratio of the lattice constants ($c/a$) with Co valence which suggests that the interlayer separation of CoO$_{2}$ sheets increases as the hole doping is varied through the optimal superconducting region.\footnote{In general there is little variation in the $a$-axis with $x$.} The origin of these structural changes is not clear at this point,  but irrespectively, these data do suggest that the electronic interlayer coupling between CoO$_{2}$ sheets becomes more 2D-like as the hole doping is varied into the optimal doping region. This trend mirrors the overall behavior of these materials in that superconductivity is achieved only for a hydrated phase with $c\sim19.8$\AA\space, while intermediate hydrate phases ($c\sim12$\AA\space),\cite{Foo} and anhydrous phases with a smaller interlayer spacing are not superconducting. We note here that the Co valence that we obtain here for superconducting compositions are similar to that of the nominally parent phase Na$_{0.7}$CoO$_{2}$ (Co$^{3.3+}$). This parent phase exhibits correlated electron behavior, and may suggest that superconductivity arises by the modification of electronic correlations that are directly present in that phase. That \Tc\space is found to be optimized for increasing interlayer separation we believe highlights the role of dimensionality and possibly a dimensionality crossover in these materials, as recently suggested by  by DFT calculations.\cite{marianetti}  Here  the hydration of  layered Na$_{x}$CoO$_{2}$ is suggested to result in a reduction in the LDA band splitting as the interlayer coupling is reduced.

Finally we suggest that it is possible that the broad plateau observed around optimal \Tc\space may result from extrinsic effects  arising from the standard method of preparation of these samples. In particular the presence of compositional gradients due to the de-intercalation of Na and intercalation of \wat\space can result in samples containing a range of \Tc 's. We believe that the relative invariance of the superconducting volume fraction in our samples over the plateau region (see above) argues  against the formation of a single point compound\cite{Jorgensen} with a single optimal \Tc\space as the origin of this plateau region.  

The measurements that we report in this letter are critical in gaining a deeper understanding of the physics of this unusual superconductor. By measuring directly the Co valence of our samples, we have established the variation of \Tc\space with electronic doping, considering both Na$^{+}$ and H$_{3}$O$^{+}$ contributions. We find a superconducting phase diagram where optimal \Tc\space is obtained for a hole doped Co$^{3+}$ system, over the wide region of Co valence states $3.24 - 3.35$, while \Tc\space decreases  for over-doped samples. These measurements suggest a band filling behavior similar to that of hole doped cuprate superconductors may also be applicable here, although the electronic configurations are quite different. The key role of dimensionality in these layered cobaltates is highlighted by the increase of the separation between CoO$_{2}$ layers, as the cobalt valence is varied into the optimal \Tc\space region. 
Our measurements show that hydrated superconductors and the anhydrous higher sodium content parent phases are similarly doped, however the change of dimensionality on hydration appears to be critical in obtaining superconductivity. This points to a strong lattice coupling to the electronic degrees of freedom and makes the understanding of the physical properties of the anhydrous compounds more pressing.

\begin{acknowledgments}
The authors thank  V. Eyert, P.G. Radaelli and L.C. Chapon for helpful discussions and comments on the manuscript. We thank K. Takada for making available to us a preprint of ref. \onlinecite{Takada2}.
\end{acknowledgments}

%\scriptsize\bibnamefont{et~al.}
%\bibliography{refs}
%

% ****** End of file template.aps ******
\end{document}